# Disorder Dynamics in Battery Nanoparticles During Phase Transitions Revealed by Operando Single-Particle Diffraction

*Jason Huang[1], Daniel Weinstock[1], Hayley Hirsh[2], Ryan Bouck[1], Minghao Zhang[2], Oleg Yu. Gorobtsov[1], Malia Okamura[5], Ross Harder[3], Wonsuk Cha[3], Jacob P. C. Ruff[4], Y. Shirley Meng[2], and Andrej Singer[1]*

[1]Department of Materials Science and Engineering, Cornell University, Ithaca, New York 14853, USA

[2]Department of NanoEngineering, University of California San Diego, La Jolla, CA, 92093 USA

[3]Advanced Photon Source, Argonne National Laboratory, Argonne, Illinois, 60439 USA

[4]Cornell High Energy Synchrotron Source, Cornell University, Ithaca, NY 14853, USA

[5]Department of Materials Science and Engineering, Carnegie Mellon University, Pittsburgh, PA 15213, USA

Structural and ion-ordering phase transitions limit the viability of sodium-ion intercalation materials in grid scale battery storage by reducing their lifetime. However, the combination of phenomena in nanoparticulate electrodes creates complex behavior that is difficult to investigate, especially on the single nanoparticle scale under operating conditions. In this work, operando single-particle x-ray diffraction (oSP-XRD) is used to observe single-particle rotation, interlayer spacing, and layer misorientation in a functional sodium-ion battery. oSP-XRD is applied to $Na_{2/3}[Ni_{1/3}Mn_{2/3}]O_2$, an archetypal P2-type sodium-ion positive electrode material with the notorious P2-O2 phase transition induced by sodium (de)intercalation. It is found that during sodium extraction, the misorientation of crystalline layers inside individual particles increases before the layers suddenly align just prior to the P2-O2 transition. The increase in the long-range order coincides with an additional voltage plateau signifying a phase transition prior to the P2-O2 transition. To explain the layer alignment, a model for the phase evolution is proposed that includes a transition from localized to correlated Jahn-Teller distortions. The model is anticipated to guide further characterization and engineering of sodium-ion intercalation materials with P2-O2 type transitions. oSP-XRD therefore opens a powerful avenue for revealing complex phase behavior in heterogeneous nanoparticulate systems.

## 1. Introduction

Sodium-ion batteries (NIBs) are emerging as a promising grid storage technology due to their low cost of energy.[1] Of particular interest are P2-type layered sodium transition metal oxide cathodes, $Na_xTMO_2$ (TM = transition metal) because of their high energy density and fast sodium-ion diffusion during cycling.[2–7] Despite their excellent kinetics, P2 type cathodes experience multiple types of phase transitions that hinder electrochemical performance. Phase transitions due to sodium ion ordering have been linked to slower diffusion of sodium ions.[7,8] The notorious P2-O2 structural phase transition combines two mechanisms that cause significant material degradation leading to poor cycle life[9]. The sliding of transition metal layers during the

P2-O2 phase transition leads to layer exfoliation after many cycles, [10,11] while the significant volume change (>20%) contributes to further structural degradation during cycling.[9,11–13] Full understanding of the mechanisms behind P2-type $Na_xTMO_2$ phase behavior remains elusive and is critical to designing durable sodium-ion batteries. [14]

In practical and functional batteries, nanoparticulates of active cathode material are surrounded by other cell components, necessitating the development of characterization techniques that can interrogate multicomponent systems. X-rays have the penetrating power that allows for operando characterization in a fully functioning cell, and X-ray powder diffraction has been used extensively to study positive electrode materials in situ.[11,15–18] For example, in-situ powder diffraction of P2-type $Na_xTMO_2$ has shown the formation of stacking faults during the P2-O2 phase transition.[11] Ex-situ x-ray powder diffraction revealed superstructure peaks corresponding to discrete $Na^+$ to $Na^+$ distances in sodium-ion ordering phases.[7] Despite its success, conventional X-ray diffraction lacks access to single particles because the signal is an average of many randomly oriented particles in the cathode typically slurry cast onto metal foils. Electron microscopy techniques can characterize single cathode nanoparticles with atomic resolution: transmission electron microscopy showed exfoliation of P2-type $Na_xTMO_2$ particles due to the P2-O2 phase transition.[10] Nevertheless, the microscopy experiments often must be done ex situ or with a specifically designed electrochemical cell that could fail to accurately mimic functional cell conditions. While P2-type $Na_xTMO_2$ are well-studied, operando measurements with single particle resolution remain elusive.

High flux synchrotron radiation and x-ray focusing allow studying single particles and grains in a variety of crystalline materials.[19–24] While some of these techniques, for example, Bragg coherent diffractive imaging (BCDI), can image single dislocations in a nanoparticle, they are only able to characterize a statistically small sample size in a reasonable amount of time.[22–25] Here, we use operando single-particle x-ray diffraction (oSP-XRD) to characterize structural dynamics of a P2-type $Na_xTMO_2$ cathode during charge. By optimizing the volume exposed to x-rays, we measure diffraction peaks from ~100 nanoparticles while maintaining single-particle resolution. The method combines the larger sample size of powder XRD with the single-particle resolution of BCDI. It allows for the operando quantification of changes in layer spacing and lattice misorientation within individual nanoparticles during the charging process in a fully operational coin cell. We study $Na_{2/3}[Ni_{1/3}Mn_{2/3}]O_2$ (NNMO) as an archetypal P2-type cathode material with a high operating voltage (~3.8 V) and specific capacity (~173 mAh $g^{-1}$).[8] NNMO exhibits both the P2-O2 phase transition and sodium-ion ordering phase transitions.[7,8,26] Capturing dynamic interactions of these phenomena will lead to better understanding of degradation at single article level and accelerate the rational design of high voltage layered materials for sodium batteries with improved electrochemical performances.

## 2. Results

Figure 1a shows the schematic of the experimental setup for the oSP-XRD. The operando coin cell contains the cathode with NNMO nanoparticles, PVDF, and acetylene black on aluminum foil, a glass fiber separator, and a sodium-metal anode. The operando system was a modified 2032 coin-cell with a 3mm and a 5mm hole upstream and downstream of the cathode to allow x-ray transmission. Both holes were sealed with polyimide tape and epoxy, which did not mitigate electrochemical performance as evidenced by the expected response of the average lattice constant of the cathode material during desodiation. The coin-cell setup is identical to systems previously used for electrochemical testing.[27,28] The average size of the cathode particles is ~500 nm. An x-ray beam with an energy of 17.1 keV and a 2-D PILATUS detector were used to collect both (002) and (004) single nanoparticle Bragg peaks while the battery was charged at a current rate of C/10 (full charge or discharge in 10 hours). An x-ray slit size of 200 μm by 60 μm was found to maximize the number of resolvable single-particle diffraction peaks. The operando coin cell sample was rocked along the scattering angle $\theta$ within the scattering plane (Figure 1a), and diffraction patterns were collected every 0.01°. Figure 1b shows a representative sample of (002) peaks collected during charge as a function of the rocking angle $\theta$ as well as $\chi$ and $2\theta$ spanned by the 2D detector. Each peak corresponds to diffraction from a single cathode nanoparticle inside the operando coin cell.

Our experiment allows us to track operando the peak position and width in all three angular directions – $2\theta$, $\theta$, and $\chi$ (see Figure 1b,c). Following Bragg's Law, the peak movement in $2\theta$ is due to the changing layer spacings (c-lattice parameter) in NNMO during sodium extraction. Most peaks moved to lower $2\theta$ angles consistent with increasing c-lattice parameter associated with sodium extraction.[7,11] In conventional powder x-ray diffraction, only the $2\theta$ peak motion and broadening is accessible due to averaging in the $\theta$ and $\chi$ directions. In oSP-XRD, we track the position and width in $\theta$ and $\chi$, enabling access to angular orientation and distortion of the measured particles similar to single crystal diffraction[29]. Particle rotation induces peak movement in $\theta$ and $\chi$, as illustrated in Figure 1c. All peaks moved in $\theta$ and $\chi$ directions indicating nanoparticle rotation by ~1° during the 10-hour charge, which was likely driven by charge transport of the particles as similarly seen in lithium-ion batteries.[28]

While observing peak position in $\theta$ and $\chi$ can inform us about particle rotation, analyzing the peak width and shape reveals disorder within single nanoparticles (see Figure 1d).[28] Following Williamson-Hall analysis, peak width broadening in $\theta$ and $\chi$ is due to size effects and misorientation,[29] and quantitative analysis for the broadening requires measuring multiple Bragg peaks from the same crystal. To measure both (002) and (004) peaks from the same set of particles, we chose rocking angle ranges of -1° to 1° for (002) and 2.75° to 4.75° for (004). As an example, Figure 1e and Figure 1f show the correlated peak movement in all three ($2\theta$, $\theta$, and $\chi$) directions confirmed that both peaks are diffracted from the same particle.

Figure 2a shows $\theta$-$\chi$ cross sections of a single selected particle's (002) peak at different compositions of $Na_x[Ni_{1/3}Mn_{2/3}]O_2$ ($0 \leq x \leq 2/3$) during charge. An overall trend of peak broadening and splitting is visible as a bright single pristine peak at x = 0.67 evolved into multiple dimmer peaks during charge. However, this trend was interrupted by two instances of peak narrowing at x = 0.51 and x = 0.36 (Figure 2a) as indicated by the emergence of one dominant peak in those frames. Nearly identical peak width behavior was also observed in the (004) peaks of the same particle in Figure S1b. Peak splitting along $\theta$ and $\chi$ occurred without noticeable splitting in the 2θ direction (Figure S2). However, high-resolution single particle NNMO (002) diffraction peaks taken at the Advanced Photon Source show peak splitting along and perpendicular to the 2θ direction around x = 0.32 (right inset of Figure 2b). The peak splitting in 2θ corresponds to a 0.01 Å decrease in layer spacing which matches well with the layer spacing change associated with the peak shift shown in Figure 3b and will be discussed later. Peak intensity is also shown to shift from on peak to another between x = 0.36 and x = 0.32 (Figure S3).

To utilize our large sample size of single-particle peaks, we calculated the average rocking curve (θ) peak width with autocorrelation. We first correlated slices of diffraction data in θ and χ, and used the half-width half maximum of the resulting peak as the average θ peak width (see Methods). The average widths of both the (002) and (004) peaks are shown in Figure 2b as a function of sodium concentration. Figure S4 shows the (002) autocorrelation θ peak width of another operando NNMO cell which showed similar peak width behavior. Since the average peak width in θ changed for both (002) and (004) peaks identically, we attribute the peak width broadening to crystal misorientation (α).[29,30] If the broadening was due to the limited crystalline size, the (004) peak would be narrower than the (002) peak in theta. We determined the degree of layer misalignment within individual crystalline grains by using the Williamson-Hall fits (see Methods). The slope of the fit is the misorientation (α) and its behavior (see left inset in Figure 2b) agreed well with the autocorrelation θ peak width behavior. In the beginning of the charge, the misorientation increased linearly from x = 0.67 to x = 0.42 as shown in red (Figure 2b). Beginning at x = 0.42, misorientation reduced from x = 0.42 to x = 0.37 which indicates layer alignment as shown in blue (Figure 2b). This alignment is followed by a faster misorientation increase until x = 0.32 as shown in green (Figure 2b). After x = 0.32, P2-NNMO is expected to undergo a transition to the sodium-free O2 phase in which the layer spacing collapses. This region is colored in yellow (Figure 2b) and misorientation gradually decreases.

To correlate layer misorientation to changes in layer spacing, the oSP-XRD data was averaged over θ and χ directions. This averaged data is equivalent to powder diffraction and is shown alongside the electrochemical voltage profile in Figure 3a. An enlarged voltage profile is shown in Figure S5. Both the powder diffraction and electrochemical data agreed well with previous studies in NNMO.[7,8,11,26,31] Lingering signal of the pristine phase persisted at all sodium concentrations indicating a small amount of electrochemically lagging nanoparticles. At x = 0.36, P2 (002) and (004) peaks reached a minimum in 2θ as shown in Figure 3b, which indicates a maximum in layer spacing (Figure S6). Both (002) and (004) peaks then shift slightly to a higher 2θ and maintain that position for the rest of charging. Figure S7 shows the differential capacity vs. voltage (dQ/dV) curve of the voltage profile shown in Figure 3a. Peaks in the dQ/dV curve for intercalation materials generally indicate order-disorder transitions or structural phase transitions[11] while valleys indicate solid solution regions. We observed valleys at 3.5 V and 4.0 V, which have been previously reported to correspond to the x = 0.50 to x = 0.33 sodium ion ordering phases along with the P2-O2 phase transition peak at 4.22 V.[7] We also observed two peaks close to one another at 3.65 V and 3.7 V, which have been previously observed but not discussed.[7,11]

## 3. Discussion

Based on the observed misorientation and layer spacing behavior we propose a model of the NNMO phase evolution mechanism (see Figure 4). The model serves to provide a plausible explanation for the evolution of misorientation in NNMO during charge and starts with the pristine P2-phase (x = 0.67) with well aligned layers (Figure 4a). Figure 4b illustrates the expansion of the c-lattice constant and the slow misorientation increase in the form of crystal mosaicity, revealed by the peak width broadening in both (002) and (004) peaks (see Figure 3) along θ, perpendicular to the scattering vector. The electrochemical profile and the dQ/dV curve (Figure 3a and Figure S7) reveal that the desodiation from $Na_{2/3}Ni_{1/3}Mn_{2/3}O_2$ occurs through a series of biphasic processes. The formation of $Na^+$-rich and $Na^+$-poor phases since the beginning of charge leads to an inhomogeneous distribution of the *c* parameter and thus the misorientation. It has been shown previously that, in NNMO, nickel is the redox-active transition metal center and donates electrons during charge.[7] As Na-ions are extracted from the cathode, nickel is oxidized from $Ni^{2+}$ in the pristine phase to $Ni^{3+}$ when moving from x = 0.67 to 0.33.[7] The electronic configuration transition from $d^8$ ($Ni^{2+}$) to $d^7$ ($Ni^{3+}$) triggers the Jahn-Teller effect in $Ni^{3+}$ octahedra and distorts TM-O layers.[32,33] Since Jahn-Teller inactive $Mn^{4+}$ and $Ni^{2+}$ separate Jahn Teller active $Ni^{3+}$ octahedra, it is plausible that distortions at lower concentrations, the slow misorientation region (Figure 2b), are localized and lead to random misorientation within transition metal oxide layers.

For the alignment region (Figure 2b in blue), we propose two possible explanations for this behavior. One possible mechanism is a disordered to ordered Jahn-Teller transition since at higher concentration of Jahn-Teller active $Ni^{3+}$ octahedra, domains of correlated Jahn-Teller distortions are preferred as predicted in $NaNiO_2$[33] and experimentally observed in $Na_{5/8}MnO_2$.[34] We therefore posit that the collinear Jahn-Teller ordering which occurs through the long-range spin interaction along the M−O−Na chain leads to a realignment of TM−O layers and a reorientation of the domains when moving from x = 0.42 to x = 0.37 (see Figure 4c). One possible ordered domain (see Figure S8) formed by tiling sextuple junctions[33] exists with a corresponding 3:1 ratio of $Ni^{3+}$ to $Ni^{2+}$ at around x = 0.42, where we observe the beginning of misorientation decrease. Another possible explanation for the observed layer alignment is a low-spin to high-spin electron configuration transition. While low-spin $Ni^{3+}$ is Jahn-Teller active due to $e_g$ orbital splitting, degeneracy in the $e_g$ orbitals of high-spin $Ni^{3+}$ eliminate Jahn-Teller distortions. Therefore, a low-spin to high-spin transition in $Ni^{3+}$ and the corresponding decrease in the Jahn-Teller distortion would result in the observed decrease in misorientation.
The subsequent increase in misorientation of the layers x = 0.37 to 0.32 (Figure 2b in green) compared to the slow misorientation region (Figure 2b in red) indicates a different misorientation mechanism proposed for the beginning of the charge. The region of faster misorientation also coincides with a decrease in lattice constant (Figure 3b) and the distinct peak splitting in the high-resolution diffraction (right inset of Figure 2b). It has reported that anionic redox is activated in this voltage region and the loss of oxygen from the TM−O layers together with the $Na^+$ removal induce the phase transition from P-type to O-type structure.[35] The formation of structurally similar P3-O3 phase boundaries has been predicted to induce low angle grain boundaries and dislocations.[36] While dislocations with Burgers vector parallel to the layers are not visible with our experimental geometry, their introduction would explain the rapid increase in misorientation until x = 0.32 (see Figure 2b). The observation of concurrent layer spacing decrease (Figure 3b), misorientation increase (Figure 2b), and peak splitting (right inset of Figure 2b) suggests that the O2 phase begins to form around x = 0.37 due to layer sliding (Figure 4c), reinforcing the hypothesis that O2-type stacking faults form in the P2 structure before the onset of the two-phase region at x < 0.33.[11] The formed domains of O2 phase maintain a similar concentration of sodium and layer spacing as the P2 phase. The transfer of intensity from one peak to another during this transition (see Figure S3) is further evidence of P2-O2 phase coexistence within single nanoparticles. Subsequent sodium extraction at x < 0.32 leads to sodium-free O2 and an associated collapse of the c-lattice constant (see Figure 4d). We associate the P2-O2 sliding transition to the 3.7 V peak and the layer exfoliation to the 4.22V peak in the differential capacity vs. voltage (dQ/dV) curve (Figure S7).

## 4. Conclusion

In this work, we demonstrate the merits of oSP-XRD as a technique that can characterize a large sample size like powder XRD with single particle resolution like that of single particle diffractive nanoimaging (BCDI). Through oSP-XRD of NNMO, we reveal the complex evolution of misorientation within individual NNMO nanoparticles during charge. Based on our x-ray data, we propose a mechanistic model for the observed misorientation and layer spacing in NNMO. Misorientation behavior suggests the emergence of $Na^+$-rich and $Na^+$-poor compositions since the beginning of charge, which is reinforced by the localized and correlated Jahn-Teller distortions with the latter facilitating the P2-O2 phase transition. Localized Jahn-Teller distortions induce random layer misorientation whereas correlated Jahn-Teller distortions align layers. Layer alignment due to the emergence of a correlated Jahn-Teller effect could therefore facilitate the P2-O2 phase transition to occur before the long 4.22 V plateau (Figure 3a and S5) associated with the layer spacing collapse during charge. Discouraging correlated Jahn-Teller effects by disrupting long range ordering could be an explanation for the effectiveness of additives at eliminating the P2-O2 transition.[37] Furthermore, the formation of intermediate phases in biphasic reactions also induces misorientations and possible fractures on the particles. While we design our model from observing the dynamic disorder, further characterization is needed to fully reveal the mechanism behind the misorientation observed. Techniques like resonant elastic x-ray scattering and scanning transmission electron microscopy have been previously used to observe Jahn-Teller induced misorientations and could be applied to systems displaying P2-O2 phase transitions.[34,38,39] Based on our proposed model, limiting long range correlated Jahn-Teller distortions could be an effective strategy to delay or eliminate the detrimental P2-O2 phase transition. Strategies such as the introduction of dopants or different ratios of transition metals, which serve to break long range order, could dramatically improve the performance of sodium transition metal layered cathodes. Finally, the demonstrated extraordinary abilities of oSP-XRD opens new avenues to study in operando systems on the single particle scale by probing structural orientation, phases, and disorder of a statistically meaningful number of single particles.

## 5. Experimental Section/Methods

*Synthesis and Cell Preparation:* NNMO was synthesized by titrating transition metal nitrates $Ni(NO_3)_2 \cdot 6H_2O$ and $Mn(NO_3)_2 \cdot 4H_2O$ into a solution of stoichiometric NaOH at a rate of 10 mL $hr^{-1}$. Co-precipitated $M(OH)_2$ was filtered with a centrifuge and washed 3 times with DI water. The dried precursors were mixed and ground with stoichiometric $Na_2CO_3$ and calcined at 500 °C for 5 hr and at 900 °C for 14 hr in air. The cathodes were assembled by mixing NNMO particles with 10 wt% acetylene black and 10 wt% PTFE and slurry cast onto aluminum foil. 1M $NaPF_6$ in propylene carbonate was the electrolyte. Glass fiber (GF/D Whatman) was used as a separator. Thin rolled sodium metal was used as the counter electrode. The operando coin cell cases had holes of 3mm and 5mm in diameter drilled in the upstream and downstream cases, respectively. The holes were sealed with polyimide tape and epoxy. The cells were assembled in an argon filled glovebox.

*Operando X-ray Experiments:* Cells were mounted on a 3D-printed sample holder and the experiment was conducted at the A2 beamline at the Cornell High Energy Synchrotron Source (Cornell University, Ithaca, NY). The cells were charged to 4.6 V at a current rate of C/10. X-ray energy was set to 17.1 keV and slits were set to 200 μm by 60 μm. X-ray data was collected using a 2D Pilatus detector to measure the (002) and (004) reflections of the cathode material. θ rocking curves of 200 points over 2° with 3 seconds per point were collected over θ ranges -1° to 1° and 2.75° to 4.75°.

High resolution diffraction experiments on were conducted at the 34 ID-C beamline of the Advanced Photon Source (Argonne National Laboratory, ANL, USA). Operando coin cells were mounted on standard sample holders manufactured using a 3D printer. A photon energy of 10 keV and sample-to-detector distance of 1.36 m were used in the experiments. Timepix (34ID) 2D detector with a pixel size of 55 μm × 55 μm was used. Rocking scans around a (002) Bragg peak, approximately 1° wide with 50–100 points and 0.5-2 second exposition, were collected.

*Analysis:* χ and 2θ angles for collected diffraction rings were calculated by fitting along circular arcs based on the diffraction and detector geometry. The centers for the circular arcs were constant for each scan which indicates consistency in the diffraction experiment and accuracy of the fitting. Diffraction patterns were averaged over the entire θ range of -1° to 1° or 2.75° to 4.75° to form the powder diffraction patterns shown in Figure 3. Slices in θ and χ with the 5 2θ values around the brightest point around both (002) and (004) peaks in 2θ were chosen for 2D autocorrelation. Autocorrelation was conducted for all time points corresponding to the same set of particles using the following equation:

$$C(k,l) = \sum_{m=0}^{M-1} \sum_{n=0}^{N-1} X(m,n) X(m-k, n-l)$$

where X is the θ by χ slice and C is the resulting autocorrelation matrix with dimensions $-(M-1) \leq k \leq M-1$ and $-(N-1) \leq l \leq N-1$. The half-width half maximum of the central 2D autocorrelation peak in the θ direction was used as the average θ peak width (β). Standard error as denoted by error bars in Figure 2b were calculated using the following equation: $SE = \frac{\sigma}{\sqrt{n}}$ where σ is the standard deviation of the 5 calculated β at each point, and n is the size of the sample (which is 5 in this case). Misorientation (α) was determined using Williamson-Hall fits by calculating the slope of the least squares regression of θ peaks widths (β) of both (002) and (004) when plotted as $\beta(q)$ vs q, where $q = \frac{4\pi}{\lambda} \sin\left(\frac{2\theta}{2}\right)$ (example fits are shown in Figure S9). The fits matched well when the estimated (000) peak was fit to the origin for all points. This indicates consistent coherence length throughout the experiment.

## Acknowledgements

We thank Dr. Long Nguyen for carefully reading and editing the paper. The work at Cornell was supported by the National Science Foundation under Award # CAREER DMR-1944907. The work at UC San Diego was supported by the National Science Foundation (NSF) under Award Number DMR1608968. Research conducted at CHESS was supported by the National Science Foundation under awards DMR-1332208 and DMR-1829070. This research used resources of the Advanced Photon Source, a U.S. Department of Energy (DOE) Office of Science User Facility, operated for the DOE Office of Science by Argonne National Laboratory under Contract No. DE-AC02- 06CH11357.

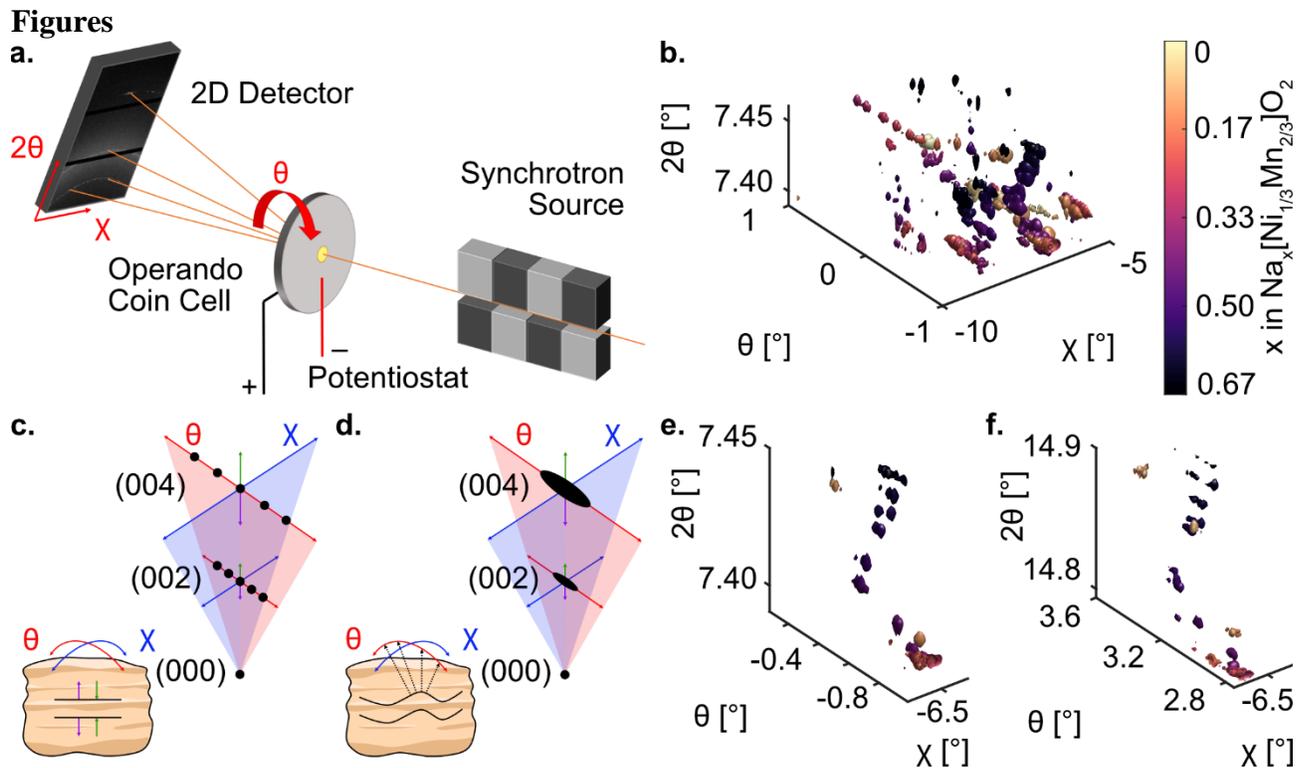

**Figure 1: Schematic of the experiment and 3D isosurface X-ray data.** (a) Experimental setup. X-ray radiation is incident on the coin cell under operando conditions. (b) An isosurface of the intensity collected around the (002) x-ray diffraction peak in θ, χ, and 2θ. (c) An illustration relating particle rotations to peak movement in θ and χ and layer spacing changes to movement in 2θ. (d) An illustration relating particle layer misorientation to peak width broadening in θ and χ. (e) Enlarged portion of (b) to highlight the evolution of a (002) peak of a single nanoparticle. (f) (004) peak of the same nanoparticle as in (e). (b), (e)-(f) use color gradient seen in color bar to indicate sodium concentration.

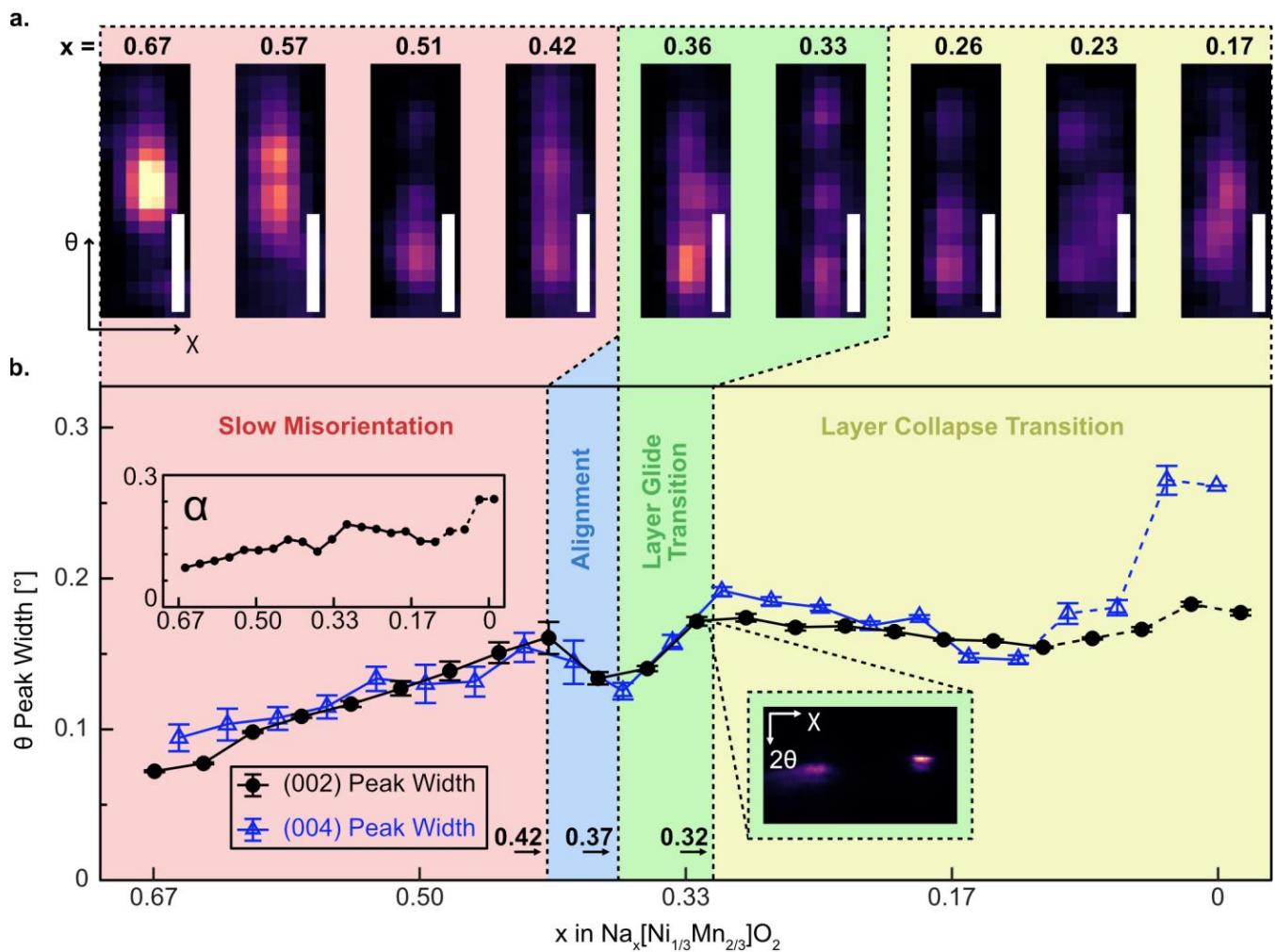

**Figure 2: Single particle peak width broadening and crystal misalignment.** (a) Enlarged oSP-XRD (002) peak intensity perpendicular to 2θ from a single NNMO particle. White scale bar indicates a 0.08° by 0.08° area. Note that the vertical aspect ratio is an artifact of higher resolution in the θ direction versus χ. Sodium concentration is indicated above each image. (b) Autocorrelation peak width along θ and calculated α misalignment in the left inset. Standard error is denoted by error bars (see Methods). At sodium concentrations of x < 0.17, (denoted by a dashed line in (b)), the peak intensity is comparable to the noise in the x-ray data. Regions of different peak width behavior are shown with different background colors. Right inset shows high resolution diffraction peak splitting during the layer glide transition region.

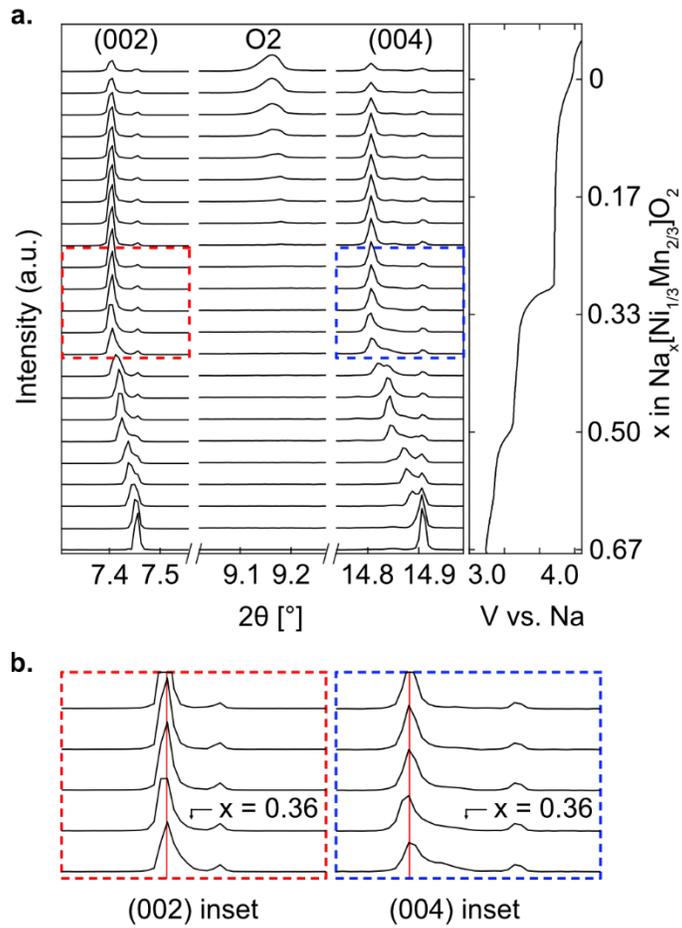

**Figure 3: Conventional powder XRD.** (a) Operando XRD intensity along 2θ for (002), and (004) peaks (λ= 0.725 Å) and voltage profile during charge at C/10. (b) Insets of (002) and (004) diffraction intensity around x = 0.33. Intensities are scaled for readability.

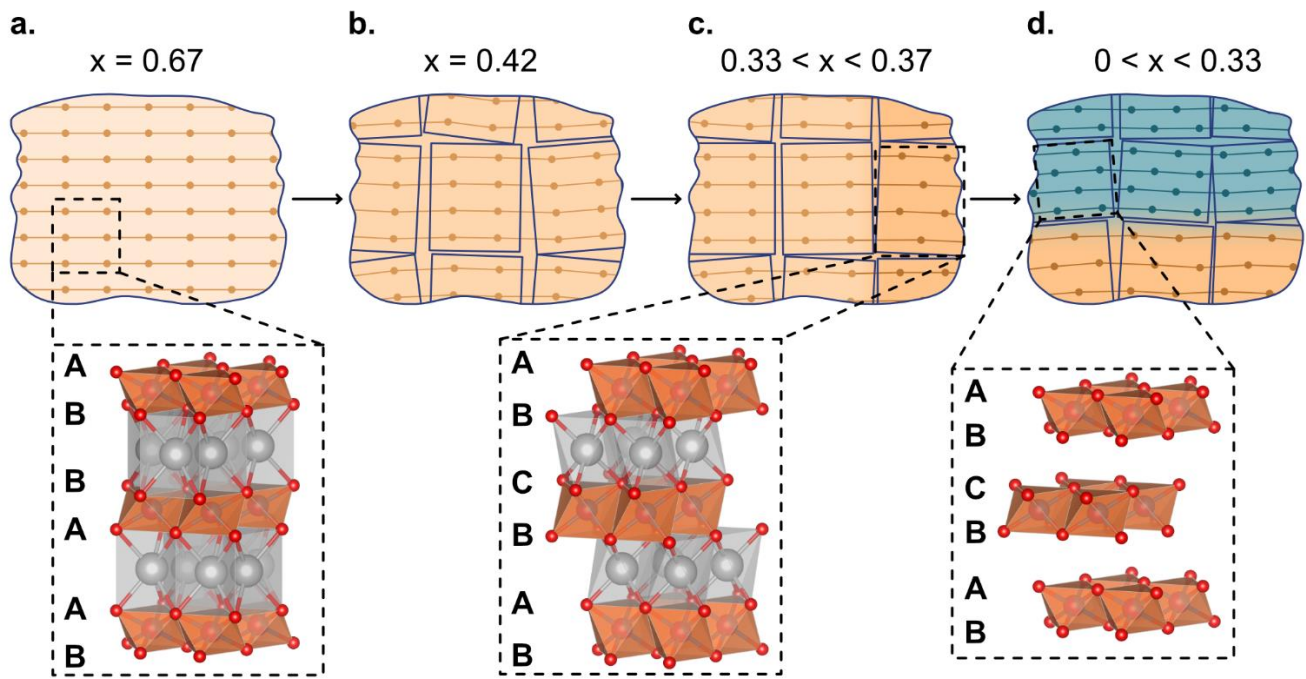

**Figure 4: Proposed model of NNMO structural phase evolution during charge and related crystal structure.** (a) Pristine P2 structure with dots representing transition metals stacked directly over each other. (b) P2 structure with expanded layer spacing and misoriented domains. (c) The P2 phase with slightly reduced misorientation is shown in light orange. The emerging O2 structure as shown in the darker orange. The P2-O2 phase boundary resides between the two domains. (d) The sodium depleted O2 phase is shown in blue with a collapsed layer spacing coexisting with the partially sodiated O2 phase in orange. Insets show visualized crystal structures of indicated phases with labeled stacking sequences of oxygen layers (TM: orange, Na: silver, O: red).

Figure S1.

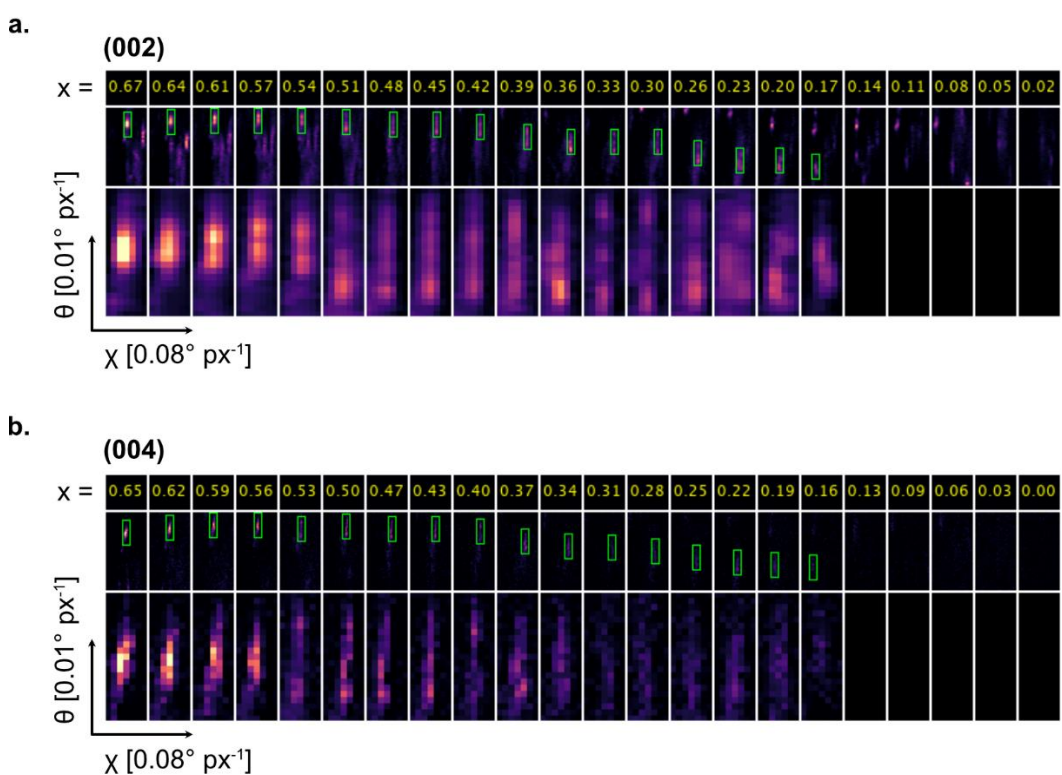

oSP-XRD peak intensities perpendicular to 2θ from a single nanoparticle (same particle as Figure S2 and Figure 3a). (a) the (002) peaks from selected nanoparticle. (b) the (004) peaks from selected nanoparticle. Green boxes indicate the location of the enlarged region. Sodium concentration for each image is shown in yellow. No enlarged region is shown when peak disappears at x < 0.14.

Figure S2.

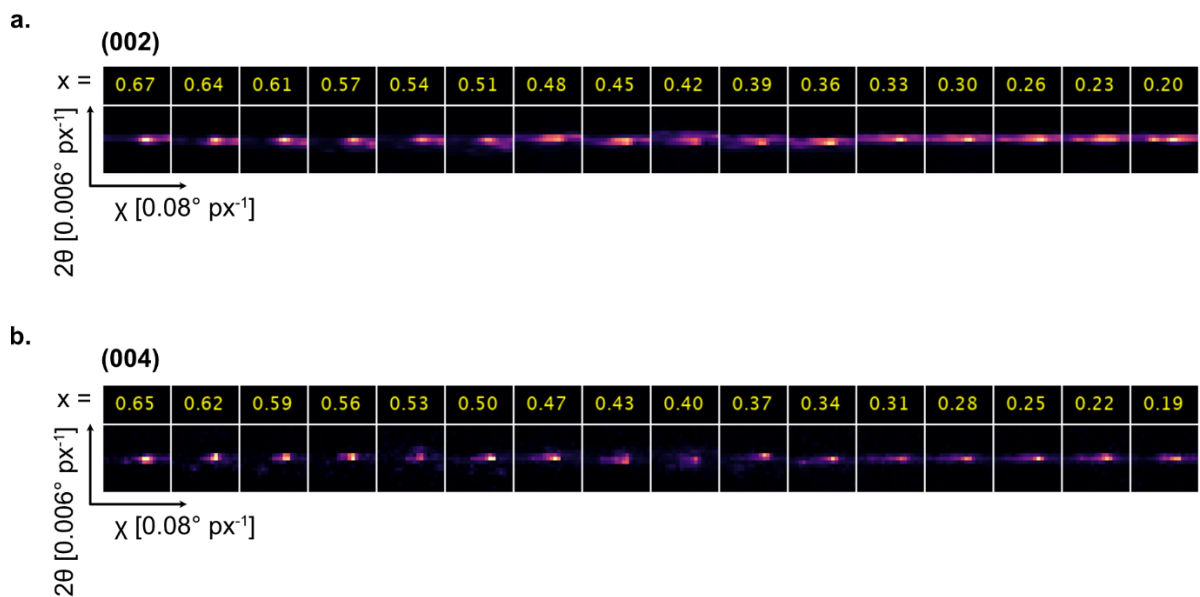

oSP-XRD peak intensities in the 2θ-χ plane from single nanoparticle (same particle as Figure S1 and Figure 3a). (a) the (002) peaks from selected nanoparticle. (b) the (004) peaks from selected nanoparticle. Sodium concentration for each image is shown in yellow.

Figure S3.

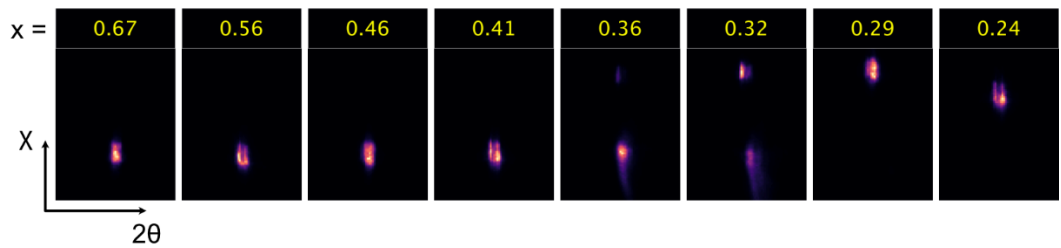

High resolution peak intensity of 2θ-χ plane of a single nanoparticle taken at APS. Sodium concentration for each image is shown in yellow.

Figure S4.

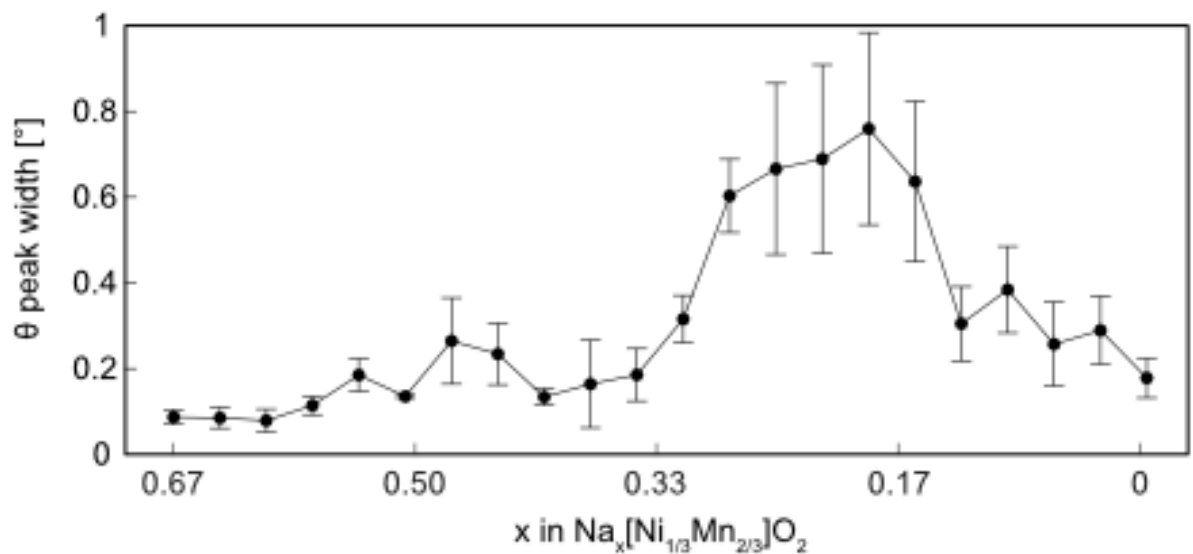

Autocorrelation θ peak width of the (002) peak from another operando NNMO cell charged at C/10. The dip observed in Figure 3b at 0.50<x<0.33 is present, and the rapid increase in misorientation around x = 0.33 is also present.

Figure S5.

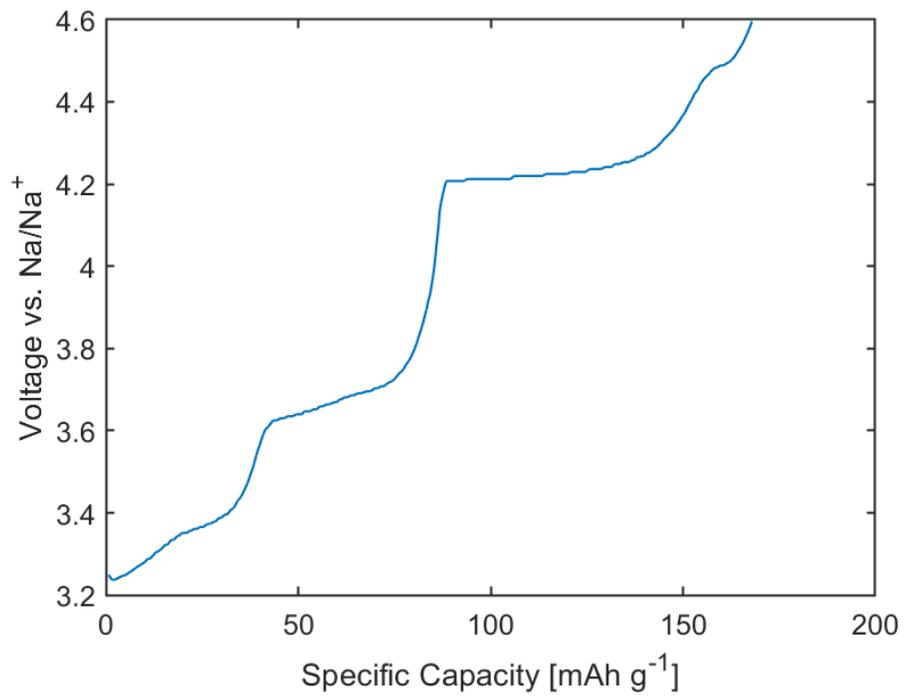

Enlarged voltage profile of NNMO operando cell charged at C/10 also shown in Figure 2a.

Figure S6.

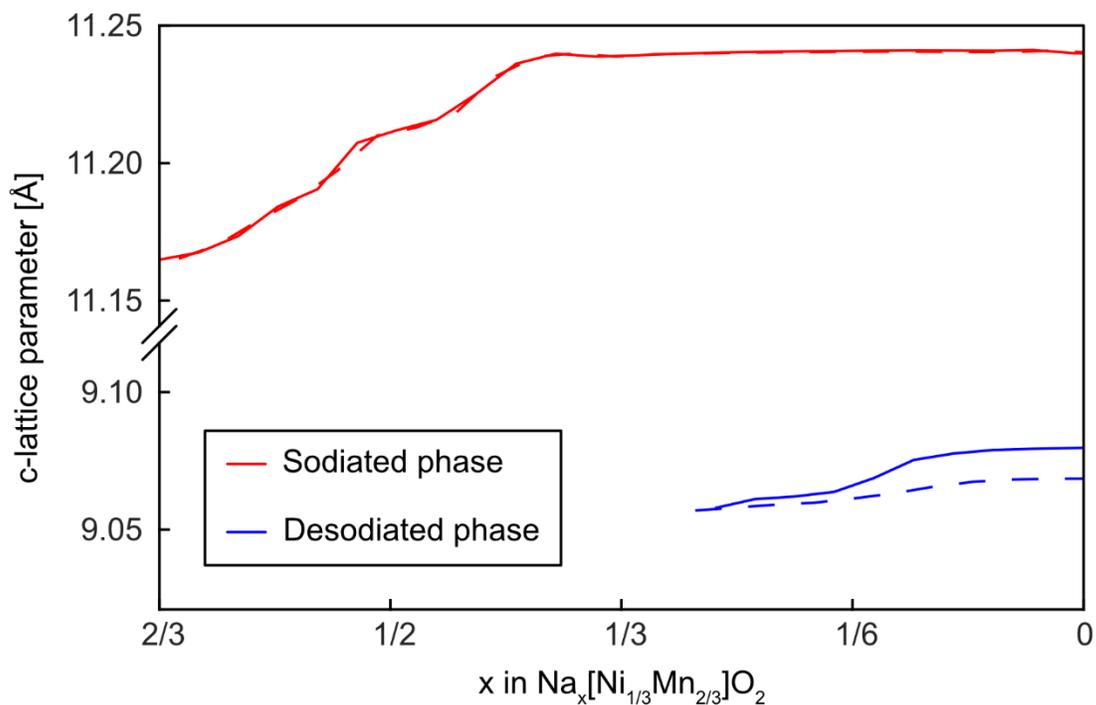

The calculated c-lattice parameter for the sodiated and desodiated phases. Dashed lines indicate different θ ranges of particles. The sodiated phase c-lattice parameter was calculated from the (002) and (004) peak positions in Figure 3, and the desodiated phase c-lattice parameter was calculated from the O2 peak positions in Figure 3. The decrease in lattice parameter highlighted in Figure 3b is also shown here.

Figure S7.

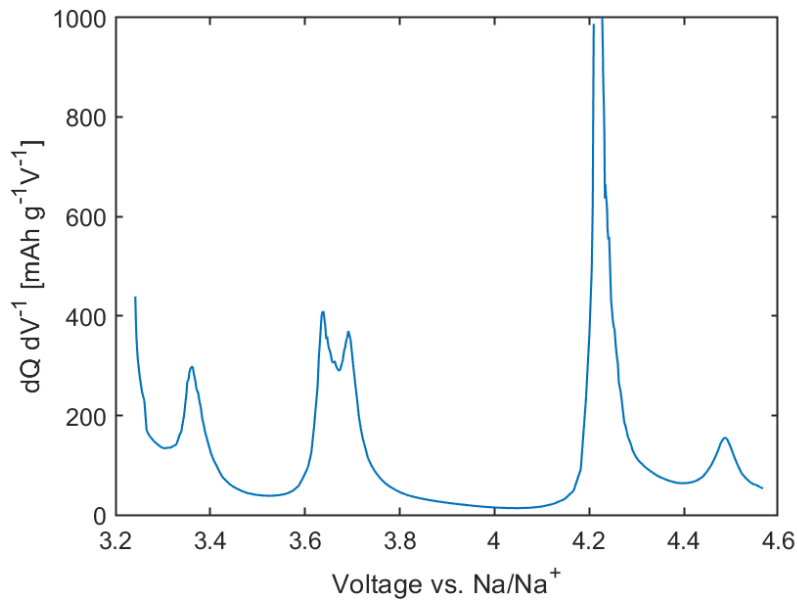

The dQ/dV profile of NNMO operando cell charged from 3.2 and 4.6V at C/10. Two intermediate phases at 3.5 and 4.0 V upon the charge were observed, which correspond to x = 0.50 and 0.33, respectively. A sharp anodic reaction was observed at 4.22 V indicating the long voltage plateau. Peaks at 3.36 V and 3.65 V correspond to sodium ion ordering transitions. Peak at 3.7 V is a result of the P2 to sodiated O2 phase transformation. Peak at 4.22V corresponds to sodiated O2 to desodiated O2 phase transition that involves large decrease in c-lattice constant and large volume change.

Figure S8.

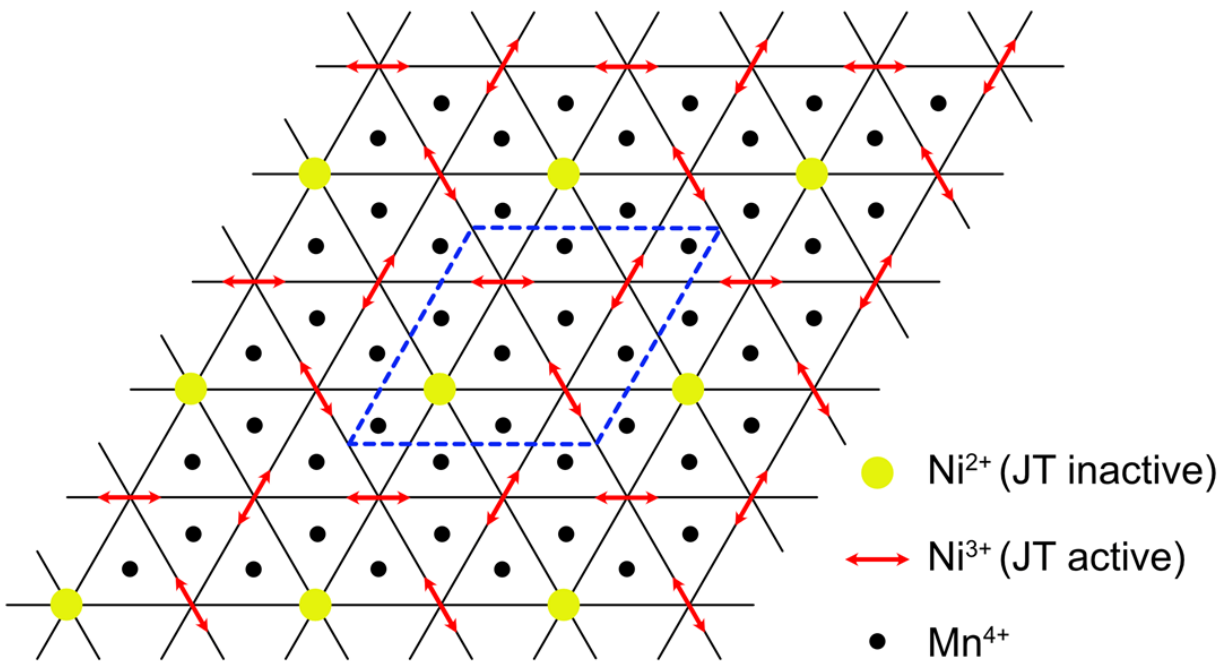

Sketch of the transition metal oxide layer projected along the [0001] direction depicting a possible Jahn-Teller ordering at a 3:1 ratio of $Ni^{3+}$ to $Ni^{2+}$. The unit cell is outlined with the dashed blue line.

Figure S9.

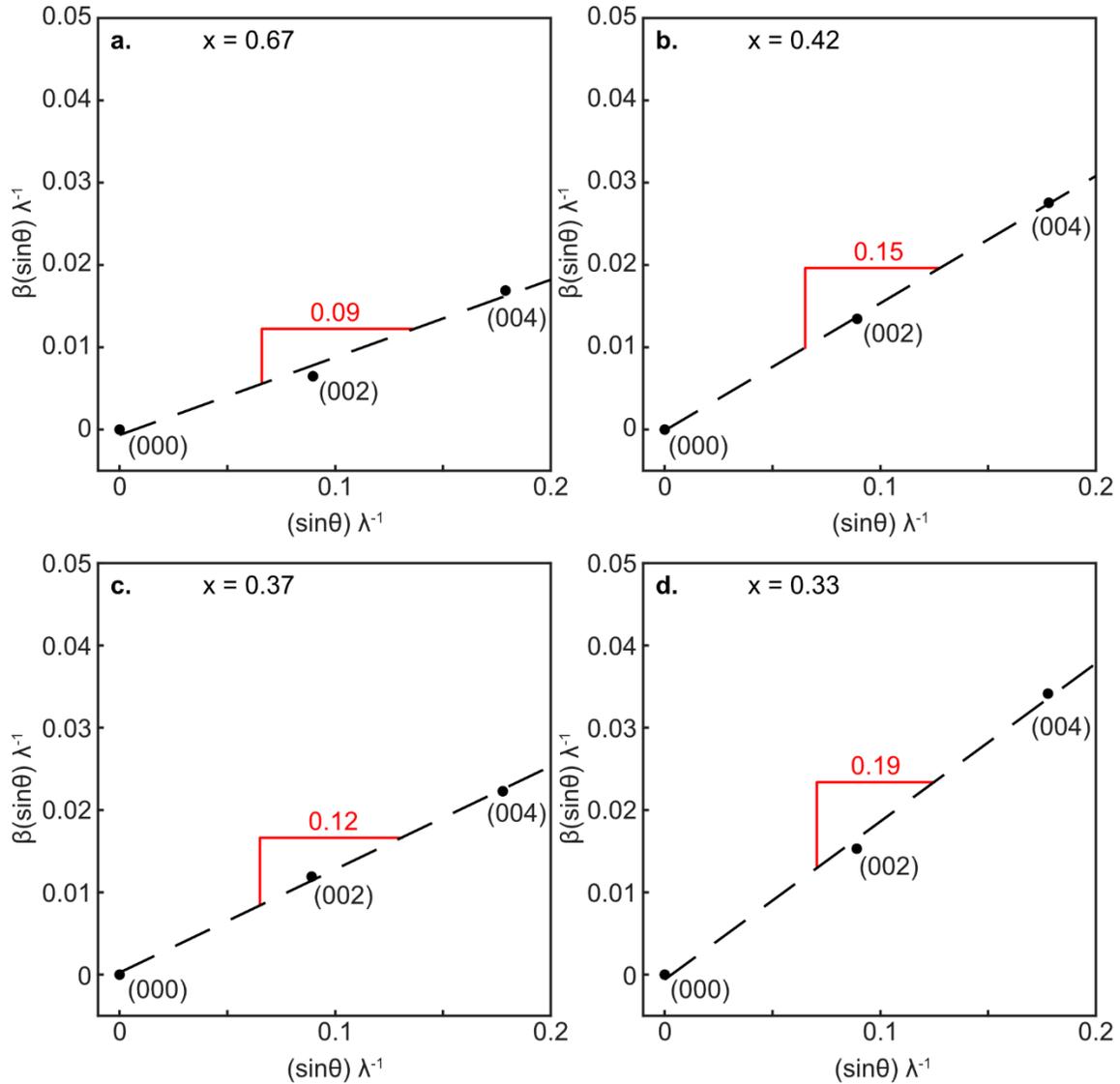

Autocorrelation θ peak width Williamson-Hall misorientation calculations via linear regression at different charge states. The θ peaks widths (β) of both (002) and (004) were used to plot $\beta(q)$ vs q, where $q = \frac{4\pi}{\lambda} \sin\left(\frac{2\theta}{2}\right)$. Points were fitted linearly with a point at the origin since the peak size broadening from a ~500 nm crystal is an order of magnitude smaller than the observed peak widths. (a) x = 0.67. (b) x = 0.42. (c) x = 0.37. (d) x = 0.33.